\documentclass[useAMS]{mn2e}
\usepackage{multirow}
\usepackage{epsfig}
\usepackage{subfigure}
\usepackage{url}

\makeatletter
\let\jnl@style=\rm
\def\ref@jnl#1{{\jnl@style#1}}
\makeatother

\title[The soft X-ray emission of Ark 120]
{The soft X-ray emission of Ark 120. XMM-$Newton$,
$NuSTAR$, and 
the importance of taking the broad view}

\author[G. Matt et al]{G. Matt$^1$, A. Marinucci$^1$, M. Guainazzi$^2$, L.W. Brenneman$^3$, M. Elvis$^3$, 
\newauthor A. Lohfink$^4$,  P. Ar\`evalo$^5$, S.E. Boggs$^6$, M. Cappi$^7$, F.E. Christensen$^8$, 
\newauthor W. W. Craig$^{8.9}$, A.C. Fabian$^{10}$, F. Fuerst$^{11}$, C.J. Hailey$^{12}$, F.A. Harrison$^{11}$, 
\newauthor M. Parker$^{10}$, C.S. Reynolds$^4$, D. Stern$^{13}$, D.J. Walton$^{11}$, 
W. W. Zhang$^{14}$\\
$^1$ Dipartimento di Matematica e Fisica, Universit\`a degli Studi Roma Tre, 
via della Vasca Navale 84, 00146 Roma, Italy\\
$^2$ European Space Astronomy Center of ESA, Apartado 50727, 28080
Madrid, Spain\\
$^3$ Harvard-Smithsonian Center for Astrophysics, 60 Garden Street,
Cambridge, MA, USA \\
$^4$ Department of Astronomy, University of Maryland, College Park, MD 20742-2421, USA \\
$^5$ Instituto de Astrof\'isica, Facultad de F\'isica, Pontificia Universidad
Cat\'olica de Chile, Casilla 306, Santiago 22, Chile \\
$^6$ Space Sciences Laboratory, University of California, Berkeley, California 94720, USA \\
$^7$ INAF, IASF Bologna, Via P Gobetti 101, 40129 Bologna, Italy\\
$^8$ DTU Space National Space Institute, Technical University of Denmark, Elektrovej 327, 
2800 Lyngby, Denmark\\
$^9$ Lawrence Livermore National Laboratory, Livermore, California 94550, USA \\
$^{10}$ Institute of Astronomy, Madingley Road, Cambridge CB3 0HA, UK \\
$^{11}$ Cahill Center for Astronomy and Astrophysics, California Institute of 
Technology, Pasadena, CA 91125, USA \\
$^{12}$ Columbia Astrophysics Laboratory, Columbia University, New York, New York 10027, US\\
$^{13}$ Jet Propulsion Laboratory, California Institute of Technology, Pasadena, CA 91109, US\\
$^{14}$ NASA Goddard Space Flight Center, Greenbelt, Maryland 20771, USA
}

\begin{document}
\maketitle
\label{firstpage}


\begin{abstract} 
 
We present simultaneous XMM-$Newton$ and $NuSTAR$ observations of the `bare'
Seyfert 1 galaxy, Ark 120, a system in which ionized absorption is absent.
 The $NuSTAR$ hard X-ray spectral coverage 
allows us to constrain different models for the excess soft X-ray emission. Among 
phenomenological models, a cutoff power law best explains the soft X-ray emission.
This model likely corresponds to Comptonization of the accretion disk seed UV photons by a 
population of warm electrons: 
using Comptonization models, a temperature of $\sim$0.3 keV and 
an optical depth of $\sim$13 are found. 
If the UV-to-X-ray {\sc optxagnf} model is applied, the UV fluxes 
from the XMM-$Newton$ Optical Monitor suggest an intermediate black hole spin. 
Contrary to several other sources observed by $NuSTAR$,
no high energy cutoff is detected, with a lower limit of 190 keV. 

\end{abstract}


\section{Introduction}

The nature of the soft X-ray excess (i.e. emission in soft
X-rays in excess of the extrapolation of the hard power law component)
has been a matter of debate since its discovery (Singh et al. 1985; Arnaud et al. 1985). The
first proposed explanation, 
pure thermal disk emission, is ruled out by the high temperature implied
(about 0.1-0.2 keV) and by its constancy over a wide range of black hole
masses (Gierlinski \& Done 2004). A great leap forward in our understanding of
the soft X-ray emission was the realization that reflection of the primary 
X-ray emission from the accretion disk can explain this excess if the disk is at least
moderately ionized (e.g. Ross \& Fabian 1993; Crummy et al. 2006; Walton et al. 2013).
However, there are cases in which even reflection may not be sufficient
(Lohfink et al. 2012) 

Ark 120 (a.k.a. Mrk 1095, $z$=0.0327) 
is an object where the soft X-ray emission 
is prominent (Vaughan et al. 2004). It is a 
Broad Line (H$\beta$ FWHM of 5800 km s$^{-1}$, Wandel et al. 1999) Seyfert 1 galaxy, 
with an estimated black hole mass of 
1.5$\times$10$^8$~M$_{\odot}$ (Peterson et al. 2004). It has been observed extensively in X-rays (see 
 Nardini et al. 2011 and references therein), always showing a strong soft excess and
never showing a Warm Absorber - it is indeed often referred to as a `bare' Seyfert 1. Nardini et
al. (2011), analysing the 2007 $Suzaku$ observation, found clear evidence for a relativistic iron
line, which was not clearly 
detected in the 2003 XMM-$Newton$ observation (Vaughan et al. 2004).

In this paper we report on simultaneous XMM-$Newton$ and $NuSTAR$ observations
of Ark 120 which demonstrate the importance of broad band observations 
to understand the soft excess in AGN. 

\section{Observations and Data reduction}

The XMM-$Newton$ observation of Ark 120 analysed in this paper started 
on 2013 February 18 
with the EPIC CCD cameras, the pn 
and the two MOS 
operated in small window and medium filter, and the RGS cameras. Source
extraction radii and screening for intervals of flaring particle background
were performed with SAS 12.0.1 
via an iterative process maximizing the Signal-to-Noise Ratio (SNR)
(see Piconcelli et al. 2004). After this process,
the net exposure time was of about 80 ks for the pn, adopting an extraction
radius of 40 arcsec and patterns 0 to 4. The pn background spectrum was
extracted from a source-free circular region with a radius of 50 arcsec.
Spectra were binned in order to over-sample the instrumental resolution by
at least a factor of 3 and to have no less than 30 counts in each
background-subtracted spectral channel. This allows the applicability of 
$\chi^2$ statistics.
The RGS spectra were reduced following the guidelines in Guainazzi \&
Bianchi (2007). 
The net exposure times are about 130 ks for RGS1 and RGS2.

$NuSTAR$ (Harrison et al. 2013) observed Ark 120 simultaneously with XMM-$Newton$ 
with its two coaligned telescopes containing Focal Plane Modules A and B (FPMA, FPMB). 
The Level 1 data products were processed with the $NuSTAR$ Data Analysis
Software (NuSTARDAS) package (v. 1.1.1). Event files (level 2 data products)
were produced, calibrated, and cleaned using standard filtering criteria
with the \textsc{nupipeline} task and the latest calibration files available
in the $NuSTAR$ calibration database (CALDB). Extraction radii for both the
source and background spectra were 1.5 arcmin. Spectra were binned in order
to over-sample the instrumental resolution by at least a factor of 2.5 and
to have a S/N ratio greater than 5 in each spectral channel. The
net exposure times are about 80 ks for both FPMA and FPMB. 

In the following, spectra are analysed with XSPEC v12.8.0. 
All errors correspond to 90\% confidence levels for one parameter of interest.
When performing joint XMM-$Newton$ and $NuSTAR$ fits we introduced a
multiplicative factor to account for differences in the absolute flux
calibrations. Fixing this factor to 1 for the pn, it is about 1.06 for
FPMA and 1.07 for FPMB (slightly depending on the model).  Similar values
are found in other AGN observed simultaneously by XMM-$Newton$ and $NuSTAR$
(Brenneman et al. 2014, Marinucci et al. 2014a,b).

\section{Spectral analysis}

\begin{figure}
\includegraphics[scale=.30]{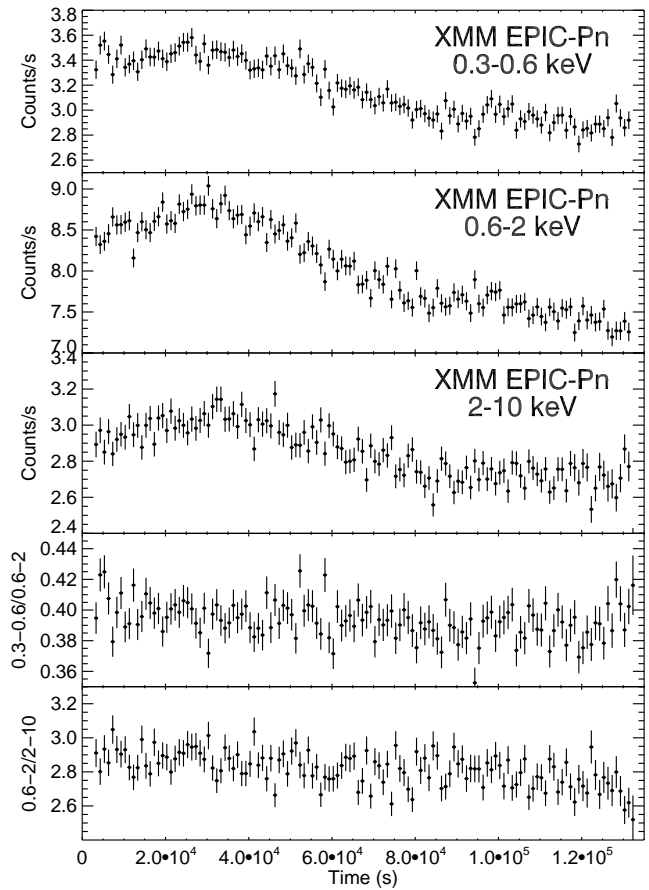}
\caption{Light curves (cts/s) of the 0.3-0.6 keV (upper panel), 0.6-2 keV (second panel from top)
and 2-10 keV fluxes (middle panel) for the XMM-$Newton$ observation (EPIC-pn).
The [0.6-2 keV]/[0.3-0.6 keV] and [2-10 keV]/[0.6-2 keV] hardness ratios are also shown (fourth and bottom
panels, respectively).
\label{hr}}
\end{figure}

\begin{figure}
\includegraphics[angle=-90,scale=.35]{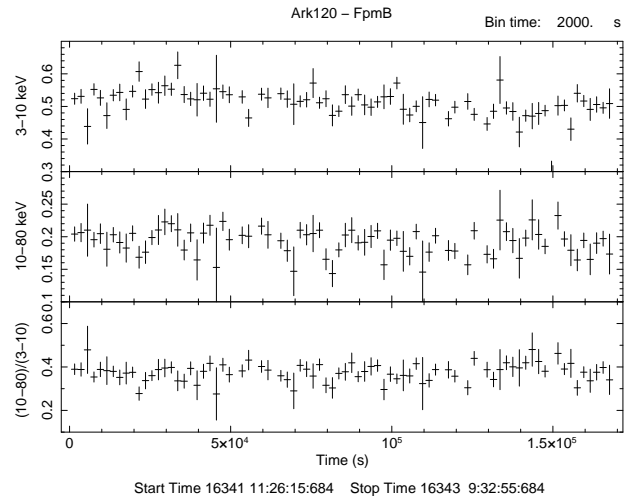}
\caption{Light curves (ct/s) of the 3-10 keV flux (upper panel), 10-80 keV flux (middle panel) and
of the hardness ratio ([10-80 keV]/[3-10 keV], lower panel) for the $NuSTAR$ observation (module B).
\label{hr_nustar}}
\end{figure}

\begin{figure}
\includegraphics[angle=-90,scale=.32]{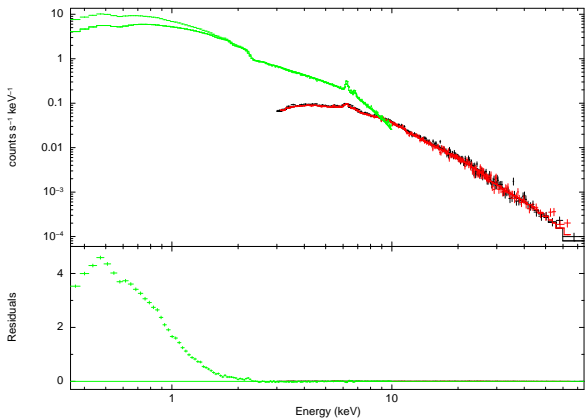}
\caption{The $NuSTAR$ 3-79 keV spectra (both modules) and best fit model extrapolated 
to the XMM-$Newton$ EPIC-pn 0.3-10 keV spectrum.
Note the strong soft excess.
\label{badfit_nustar}}
\end{figure}

\begin{figure}
\includegraphics[angle=-90,scale=.32]{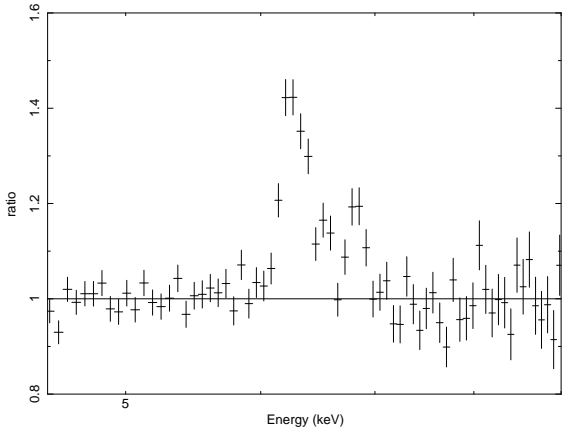}
\caption{Data to model ratio for the XMM-$Newton$ EPIC-pn spectrum in the iron line region when a
power law in the 3-5 and 7.5-10 keV ranges is fitted.
\label{ironline}}
\end{figure}

The XMM-$Newton$ 0.3-0.6 keV, 0.6-2 keV and 2-10 keV light curves show flux variations of about 15\%,
with only a small (less than 10\%) variation
in the hardness ratios. Similar flux variations, and no spectral variability,
is found in the $NuSTAR$ data. Therefore, for both satellites we used the spectra 
integrated over the entire observations and fitted them together.

First, however, we fitted the two datasets independently. 
A good fit to the 3-79 keV $NuSTAR$ spectrum is found with a relatively simple model:
a power law with a high energy cutoff ($\Gamma$=1.79$\pm$0.03, $E_c >$340 keV), 
a cold reflection model ({\sc pexrav} in XSPEC;  $R$=0.26$\pm0.08$), and
two narrow iron lines (energies of 6.41$^{+0.03}_{-0.01}$ keV and 6.91$\pm0.08$ keV). 
The $\chi^2$ is 343.0 for 343 d.o.f. ($\chi^2_r$=1). 
Then the XMM-$Newton$ spectrum was added, to show the presence of a strong 
soft excess below 2 keV (Fig.~\ref{badfit_nustar}). 

Then, we fitted the XMM-$Newton$/EPIC-pn spectrum alone in the 0.3-10 keV band
(ignoring the 1.8-2.5
keV band to exclude the instrumental features present there). For simplicity, we used the
EPIC-pn detector only. We used a model composed of
a power law absorbed by both the Galactic column ($N_{\rm H}$=9.78$\times$10$^{20}$ cm$^{-2}$,
Kalberla et al. 2005) and an intrinsic absorber (which, however, in all fits is found 
to be very small or
negligible), neutral and ionized reflection components ({\sc xillver} in XSPEC, Garcia et al. 2013),
and iron emission lines in addition to those present in the reflection models to account
for further features present in the 6.4-7 keV region (see Fig.~\ref{ironline}). 
A single reflector is clearly insufficient to fit the spectrum 
($\chi^2_r$/d.o.f.=8.51/150 and 3.38/147 for an
unblurred and a relativistically blurred reflector, respectively), and 
even with two unblurred plus two relativistically
blurred reflection components the fit is unacceptable ($\chi^2_r$/d.o.f.=2.45/141), clearly requiring
a further component to model the soft X-ray emission. The inclusion of a black body emission
component improves the fit significantly ($\chi^2_r$/d.o.f.=1.54/139), 
but the temperature found is quite large for an 
AGN (about 0.1 keV, while for a standard optically thick, geometrically thin disc
the maximum temperature appropriate for the black hole mass of Ark 120 is about 10-20 eV, 
Frank et al. 2002). A 
multicolour black body model ({\sc diskbb} model in XSPEC) results in a
fit of similar quality ($\chi^2_r$/d.o.f.=1.52/139). 

A better fit is found substituting the black body component with 
a second, steep power law ($\Gamma_{\rm Soft}\sim$2.4, $\chi^2_r$/d.o.f.=1.43/139),
even if at the expense of a very flat hard power law ($\Gamma_{\rm Hard}\sim$1.2). 
A significantly better fit ($\chi^2_r$/d.o.f.=1.29/145) is found  
with a cutoff power for the soft excess plus the primary power law, 
two unblurred reflection components and the two narrow lines; 
the addition of a blurred reflection component does
not improve the fit quality. Even if the $\chi^2$ for these fits is not ideal,
an inspection of the residuals shows that there are no obvious features left;
most of the remaining problems are related to a still imperfect fitting of the iron line region.
Because we are interested here mainly in the soft X-ray emission, we decided to accept
these fits. A detailed discussion of the reflection and line features is deferred to
a future paper, that we plan to write after
the public release of an improved energy scale calibration affecting recent
EPIC-pn observations (see discussion in Marinucci et al. 2014b). 
Here we just remark that the use of a different reflection model,
namely the {\sc reflionx} model in XSPEC (Ross \& Fabian 2005), gives similar parameters
and does not significantly affect the results on the soft excess. 

No ionized absorption is apparent, either in the EPIC or in the RGS data, confirming
the `bare' Seyfert 1 nature of Ark~120. 
The observed 0.5-2 keV and 2-10 keV fluxes are 1.4$\times$10$^{-11}$ erg cm$^{-2}$ s$^{-1}$ and
2.3$\times$10$^{-11}$ erg cm$^{-2}$ s$^{-1}$, respectively, corresponding to absorption-corrected
luminosities in the same bands of 
3.4$\times$10$^{43}$ erg s$^{-1}$ and 5.6$\times$10$^{43}$ erg s$^{-1}$.
The source is about 1.4 times fainter than in the 2007 $Suzaku$ observation, 
and 1.8 times fainter than in the 2003 XMM-$Newton$ observation. 

Even if Ark 120 is a radio-quiet source, it is not radio-silent, so we also tried
the {\sc srcut} model in XSPEC, which describes the synchrotron spectrum from an 
exponentially cut off power law distribution of electrons in a homogeneous magnetic field. 
The spectrum is similar to a cutoff power law, but with a rollover slower than exponential.
A good fit ($\chi^2_r$/d.o.f.=1.31/145) is found. One of the model's parameters
is the 1 GHz flux, which is well below the observed 1.4 GHz value (e.g. Condon et al. 1998), but
this is because the best fit power law index is almost zero. If the radio index observed for this 
source, 0.6 (Barvainis et al.
1996), is adopted, the fit is significantly worse ($\chi^2_r$/d.o.f.=1.57/146)
and the 1.4 GHz flux is almost 3 orders of magnitude larger than observed.  

A decent fit ($\chi^2_r$/d.o.f.=1.46/146) is found with a thermal bremsstrahlung model
instead of the cutoff power law. However, assuming that the emitting region
is optically thin to Thomson scattering, a lower limit to the size of the emitting region of about 0.1 pc
is found, inconsistent with the observed X-ray variability below 0.6 keV, where the soft excess component
dominates.

\begin{figure}
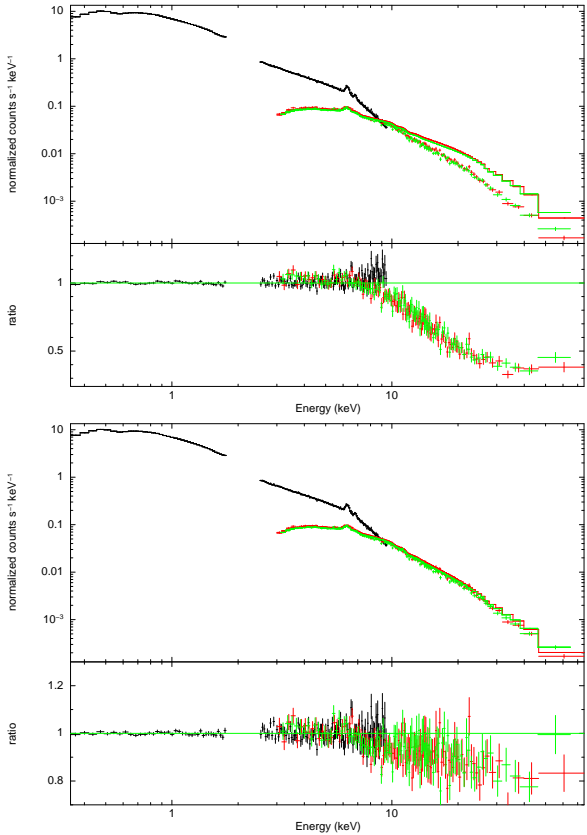

\includegraphics[angle=-90,scale=.32]{pow_reb_ext.ps}
\hfill
\includegraphics[angle=-90,scale=.32]{cutoffpl_reb_ext.ps}
\caption{{\it Upper panel}: The XMM-$Newton$ 0.3-10 keV spectrum and best fit model extrapolated to the 
$NuSTAR$ 3-79 keV spectra.
Here, the model is composed of a power law, reflection components, Gaussian iron lines, and a 
second, steeper power law to reproduce the soft X-ray excess. {\it Lower panel:}  The same, 
but with a cutoff power law for the soft excess.
\label{powextr}}
\end{figure}

To summarize, from the XMM-$Newton$ data alone we found that a cutoff power law for the 
soft X-ray excess is preferred to, e.g., a power law, but the statistical
difference is such that the latter model cannot be entirely ruled out. 
However, the issue becomes immediately clear once we extrapolate (without refitting) 
the XMM-$Newton$ best fit models to the $NuSTAR$ band (see Fig.~\ref{powextr}): 
the power law model completely fails to fit the $NuSTAR$ spectra. 

\begin{figure}
\includegraphics[angle=-90,scale=.32]{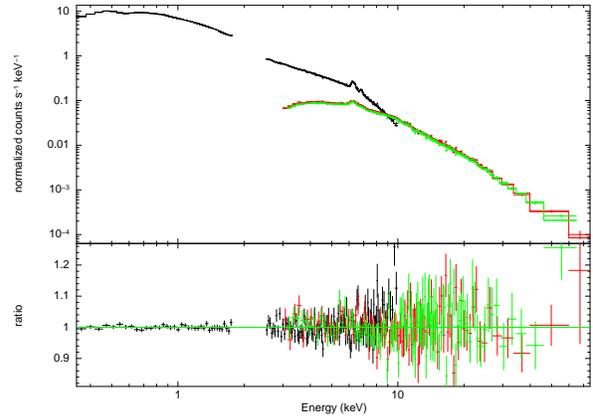}
\caption{Combined XMM-$Newton$ and $NuSTAR$ spectra and best fit model, when a cutoff power law
is adopted for the soft excess.
\label{bestfit}}
\end{figure}

This result is confirmed by the joint XMM-$Newton$/$NuSTAR$ analysis. 
The cutoff power law modeling of the soft X-ray emission (i.e. the refitting
of the model presented in the lower panel of Fig.~\ref{powextr}) provides by far 
the best fit ($\chi^2_r$/d.o.f.=1.14/494, see also Fig.~\ref{bestfit}) 
while the next best fit, that with a black body, has $\chi^2_r$/d.o.f.=1.36/488.

\begin{figure}
\includegraphics[angle=-90,scale=.32]{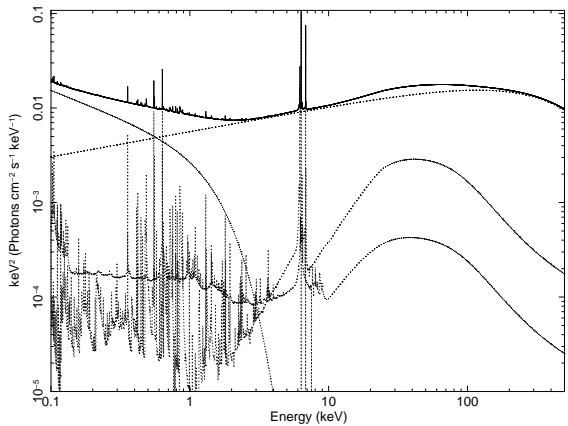}
\caption{The best fit model, with all components also separately shown, 
for the combined XMM-$Newton$ and $NuSTAR$ spectra when the {\sc optxagnf} model is adopted
for the soft excess.
\label{bestfitmodel}}
\end{figure}

Very similar quality of fits are found using, for the soft excess, 
Comptonization models instead
of a cutoff power law (e.g. {\sc compTT}, {\sc nthcomp} or {\sc optxagn} in XSPEC). 
All these models give very similar values for the temperature and optical depth 
of the Comptonizing slab. We
discuss here in some detail the {\sc optxagnf} model (Done et al. 2012). In this model 
the gravitational energy released in the disc 
at each radius is emitted as a (colour temperature corrected) blackbody only down to 
a given radius, the coronal radius $R_c$. Below this radius, it is assumed that the energy 
can no longer completely thermalize, and is distributed between a low energy and a high
energy electron population, giving rise to a soft and a hard Comptonization components.  
Even if some of the model assumptions (e.g. the sharp edge between the thermalization and
scattering regions, the geometrical coincidence of the two scattering zones, 
the maximum disk temperature fixed to that at the coronal radius)
are certainly oversimplified, it has the merit of connecting the UV and soft X-ray emission.
Input parameters of the model are the black hole mass and the distance of the source (which
we kept fixed to the values mentioned in the introduction), the black hole spin,
the bolometric Eddington ratio $L/L_{Edd}$, 
$R_c$, the electron temperature and optical depth of the low energy electron population,
the power law index of the high energy emission and the fraction, $f$, of the power below 
the coronal radius which is emitted in the hard comptonization component. Because
in {\sc optxagn} the hard component is parametrized as a power law with a cutoff energy of 100 keV,
we first estimated $f\sim0.67$ using XMM-$Newton$ data alone and then switched off the hard component,
substituting it with a cutoff power law. 
The spin could not be constrained, so we initially fixed it to zero. 
The reduced $\chi^2$ is 1.14 for 493 d.o.f.. The best fit parameters are listed in Table~\ref{optxagn}
while the best fit model (without Galactic absorption for clarity) is shown 
in Fig.~\ref{bestfitmodel}. Besides
{\sc optxagn} and the cutoff power law, the model is composed of two reflection components and
two Gaussian iron lines. 
Though a detailed discussion of the reflection and line emission is deferred to a future
paper, we can say here that the best fit is obtained with two reflection components with ionization 
parameters of $\sim$50 (the brightest) and $\sim$1000 erg cm s$^{-1}$, 
respectively.  The iron abundance is $\sim$3.4. Two emission
lines are also needed, one at about 7 keV (possibly related to Fe XXVI;  the iron K$\beta$
line is already included in the {\sc xillver} reflection model), with a significance
of $>$99.99\%  according to an F-test, the other at about 6.55
(corresponding to Fe XXI-XXII), with a significance of 99.98\%. 

We then fixed the spin to 0.5 and to 0.99. The best fit parameters
are listed in Table~\ref{optxagn}. It is worth noting that when we increase the spin,
the larger emitting area due to the lower value of the
innermost stable orbit is compensated by a lower $L/L_{Edd}$ ratio. The two parameters
are therefore largely degenerate, and from the X-ray spectra alone it is not possible 
to measure the black hole spin.

\begin{figure}
\includegraphics[angle=-90,scale=.32]{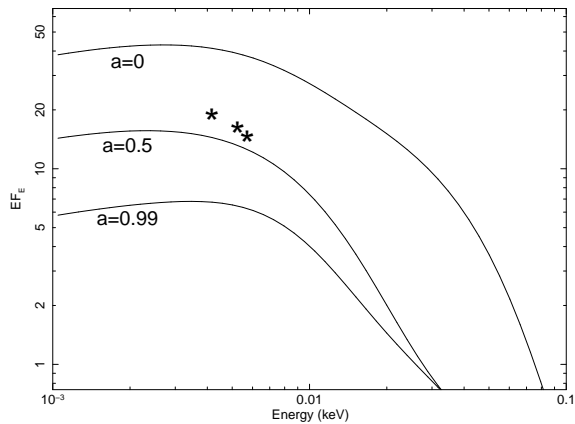}
\caption{Extrapolation of the combined XMM-$Newton$ (EPIC-pn) and $NuSTAR$ best fit models
for $a$=0, 0.5 and 1 to the XMM-$Newton$ OM UV fluxes. 
\label{UV}}
\end{figure}

To remove this degeneracy, we used the XMM-$Newton$ Optical Monitor data, similarly to what
was done by Done et al. (2013).
Ark 120 was observed with filters UVW1, UVM2 and
UVW2, whose effective wavelengths are 2910, 2310 and 2120 \AA, respectively. The fluxes
were 3.9$\times$10$^{-14}$, 5.3$\times$10$^{-14}$ and 5.6$\times$10$^{-14}$ erg cm$^{-2}$ s$^{-1}$ 
\AA$^{-1}$ (with statistical errors of about 1\%), respectively, 
after correction for Galactic extinction 
(Schlafly \& Finkbeiner 2011) following Seaton (1979). Contributions from the host galaxy are estimated
to be less than 10\%. Given the uncertainties in any extinction
correction and the simplifying assumptions in the model (including the fact that it is angle-averaged,
so corresponding to an inclination angle of 60 degrees; for lesser inclination angles the flux 
would be higher, up to twice as much for a face-on disc), we did not attempt to
fit the UV-X-rays SED but limited ourselves to extrapolating the best fit model and comparing
it to the UV fluxes. The results are shown in Fig.~\ref{UV}. The $a$=0.99 model falls short of the
UV fluxes by a factor $\sim$2-3, while the $a$=0 is larger by a factor $\sim$2. 
The extrapolation of the $a$=0.50 model, instead, 
is roughly consistent with the UV fluxes. While the range of  $L/L_{Edd}$ values are rather common
for bright Seyfert galaxies (e.g. Steinhardt \& Elvis 2010), 
it is interesting to note that the low spin solution
gives a $L/L_{Edd}$ ratio larger than the typical values 
for this source (e.g. Woo \& Urry 2002, Peterson et al. 2004),
an unlikely situation given that in our observation the source was in a rather low flux state. 
We conclude that an intermediate value for the spin is preferred. Interestingly, 
Nardini et al. (2011) reached the same conclusion based on a relativistic reflection
fit (see also Patrick et al. 2011). 

\begin{table}
\begin{center}
\caption{Best fit parameters for the joint XMM-$Newton$/$NuSTAR$ fit
with the soft excess modeled by {\sc optxagnf} (Done et al. 2012), for
three different black hole spins. \label{optxagn}}
\begin{tabular}{cccc}
\hline\hline
$a$ & 0 & 0.50 & 0.99\\
$L/L_{Edd}$ & 0.16$^{+0.16}_{-0.08}$ & 0.05$^{+0.01}_{-0.01}$ & 0.04$^{+0.03}_{-0.01}$\\
$R_c$ ($R_G$) & 11.5$^{+0.1}_{-3.4}$  & 31.3$^{+39.2}_{-16.6}$ & 24.9$^{+16.0}_{-15.2}$ \\
$kT$ (keV) & 0.33$^{+0.02}_{-0.02}$ & 0.32$^{+0.01}_{-0.01}$ & 0.32$^{+0.02}_{-0.01}$ \\
$\tau$ & 12.9$^{+1.1}_{-0.9}$ & 13.6$^{+0.6}_{-0.2}$ & 13.6$^{+0.4}_{-0.7}$\\
$\Gamma$ & 1.73$^{+0.02}_{-0.02}$ & 1.73$^{+0.02}_{-0.02}$ & 1.73$^{+0.02}_{-0.02}$ \\
$E_c$ (keV) & $>$190 &  $>$190 & $>$190 \\
\hline
\end{tabular}
\end{center}
\end{table}

Because the hard X-ray emission is thought to originate from the Comptonization of UV/soft
X-ray photons by a population of hot electrons, we
substituted the hard cutoff power law with a Comptonization model. We used the
{\sc compPS} model in XSPEC, assuming a spherical geometry, a Maxwellian electron distribution
and a black body input temperature of 20 eV. The fit is as good as the one with the 
cutoff power law, and  
the parameters are not very well determined, not surprising given that
the cutoff power law model gives only a lower limit to the high
energy cutoff. For $a$=0 the electron temperature is between 110 and 210 keV, while the
optical depth is between 0.6 and 1.5. Similar values are found for the other values of the spin.
Therefore, the electron distribution is consistent with being
optically thick, which
may explain the lack of evidence for relativistic reflection: if the corona is extended (as assumed
in the {\sc optxagnf} model), reflection from the inner disk may be scattered and then 
unrecognizable (e.g. Petrucci et al. 2001).

\section{Discussion and Conclusions}

The 0.3-80 keV spectral energy range provided by the quasi-simultaneous XMM-$Newton$/$NuSTAR$
observations of the `bare' Seyfert 1 galaxy Ark 120 has allowed us to study 
the soft X-ray excess in this source with 
unprecedented accuracy and robustness. Differently from many other Broad Line Seyfert 1s,
where ionized reflection is sufficient to explain the soft X-ray emission, 
a further component is required. While Ark 120 may well be a peculiar source, it should
also be noted that it is one of the best sources where to search for such a component given
the lack of intrinsic absorption. A cutoff power law is the best phenomenological
model for the excess. If interpreted as a signature of Comptonization,
a temperature of $\sim$0.3 keV and an optical depth of $\sim$13 are required. A
word of caution is needed here, as such large optical depths can work efficiently
in Comptonizing photons only if the absorption opacity is much lower than the
scattering opacity (see discussion in Done et al. 2012), and it is likely that
the presently available Comptonization model are still too simplistic, expecially
in this regime. Even with these limitations in mind, adopting
the recently developed {\sc optxagnf} model (Done et al. 2012), and using the X-ray
data alone, we find that some of the parameters, and in particular the black hole spin
and the $L/L_{Edd}$ ratio, are highly degenerate, and we find no strong  
constraint on the black hole spin from the X-ray data alone. 
The extrapolation of the best fit models to the UV fluxes, however, suggests
that an intermediate black hole spin solution is preferred. 

X-ray variability is very similar at all energies (Fig.~\ref{hr}). In the Done et al. (2012)
scenario, this means that what is driving variability is a change in the total power (possibly
related to changes in the accretion rate) rather than a change in the relative fraction of power
in the hard and soft components.

Unlike other AGN observed by $NuSTAR$ (e.g. IC4329A, Brenneman et al. 2014; MCG--5-23-16, Harrison
et al, in prep.; Swift J2127.4+5654, Marinucci et al., 2014a), 
no high energy cutoff is detected, with a lower limit to the $e$-folding energy of 190 keV.
Applying Comptonization models, a temperature of the Comptonizing region much higher
than observed in the abovementioned sources is found. 

No clear evidence for relativistic reflection is found. It is interesting to note that
in other sources observed simultaneously by XMM-$Newton$ and $NuSTAR$, where the
relativistic reflection is clearly present, this component entirely accounts for the
soft X-ray emission (MCG--6-30-15, Marinucci et al., 2014b; Swift J2127.4+5654, Marinucci et
al., 2014a, NGC 1365, Walton et al. 2014). One possible explanation
is that in Ark 120 the hot corona, responsible for the hard X-ray emission, is both 
optically thick and
extended (a possibility which is indeed consistent with the best fit value of the radius
of the Comptonizing region found with the {\sc optxagnf} model). 
The presence of a relativistic iron line, as well as a larger
reflection component in the $Suzaku$ observation,
may then be explained in terms of a less thick and/or more compact corona during that
observation. While a detailed comparison
of the present observations with previous ones is beyond the scope of this paper, we note that the
spectrum in the $Suzaku$ observation was significantly steeper, suggesting an optically thin corona.
In the future, it would be interesting
to search for correlations between the coronal parameters and the presence and strength of the
relativistic reflection in a large sample of objects, a task requiring sensitive, broad band
observations like the one described in this paper. 

\section*{Acknowledgements}
We thank the anonymous referee for useful comments which helped
improve the clarity of the paper, and Chris Done for comments and advices
on the {\sc optxagnf} model.
This work has made use of data from the $NuSTAR$ mission,
a project led by the California Institute of Technology,
managed by the Jet Propulsion Laboratory, and funded by the
National Aeronautics and Space Administration. We thank
the $NuSTAR$ Operations, Software and Calibration teams for
support with the execution and analysis of these observations.
This research has made use of the $NuSTAR$ Data Analysis Software (NuSTARDAS) jointly
developed by the ASI Science Data Center (ASDC, Italy) and the California Institute of
Technology (USA). The work is also 
based on observations obtained with XMM--$Newton$, an ESA science mission with 
instruments and contributions directly funded by ESA Member States and the USA
(NASA). GM and AM acknowledge financial support from Italian Space Agency under grant 
ASI/INAF I/037/12/0-011/13 and from the European
Union Seventh Framework Programme (FP7/2007-2013) under grant agreement
n.312789.

\end{document}